\documentclass[twocolumn]{aastex61}
\begin{document}
\newcommand{\msun}{$M_\sun$}
\newcommand{\feh}{[Fe/H]}
\newcommand{\kms}{km s$^{-1}$~}
\newcommand{\ug}{{\it ugriz}~}
\newcommand{\ugp}{$u'g'r'i'z'$~}
\newcommand{\gi}{$ (g-i)~$}
\newcommand{\hi}{H{\footnotesize I}}
\defcitealias{adams16}{A16}
\setcitestyle{notesep={; }}

\shorttitle{Detection of an Optical Counterpart to AGC~249525}
\shortauthors{Janesh et al.}

\title{Detection of an Optical Counterpart to the ALFALFA Ultra-compact High Velocity Cloud AGC~249525} 

\author{William Janesh} 
  \affiliation{
  Department of Astronomy, Indiana University, 727 E. Third Street,
  Bloomington, IN 47405, USA}

\author{Katherine L. Rhode}
  \affiliation{
  Department of Astronomy, Indiana University, 727 E. Third Street,
  Bloomington, IN 47405, USA}

\author{John J. Salzer}
  \affiliation{
  Department of Astronomy, Indiana University, 727 E. Third Street,
  Bloomington, IN 47405, USA}

\author{Steven Janowiecki}
  \affiliation{
  International Centre for Radio Astronomy Research,
  University of Western Australia,
  35 Stirling Highway,
  Crawley, WA 6009, Australia.}
  
\author{Elizabeth A. K. Adams}
  \affiliation{
  Netherlands Institute for Radio Astronomy (ASTRON), Postbus 2, 7900 AA
  Dwingeloo, The Netherlands }
  
\author{Martha P. Haynes}
  \affiliation{
  Center for Radiophysics and Space Research, Space Sciences Building, 
  Cornell University, Ithaca, NY 14853, USA }
  
\author{Riccardo Giovanelli}
  \affiliation{
  Center for Radiophysics and Space Research, Space Sciences Building, 
  Cornell University, Ithaca, NY 14853, USA }
  
\author{John M. Cannon}
  \affiliation{
  Department of Physics and Astronomy, Macalester College, 
  1600 Grand Avenue, Saint Paul, MN 55105, USA }

\begin{abstract}
We report on the detection at $>$98\% confidence of an optical
counterpart to AGC 249525, an Ultra-Compact High Velocity Cloud
(UCHVC) discovered by the ALFALFA blind neutral hydrogen survey.
UCHVCs are compact, isolated \hi\ clouds with properties consistent
with their being nearby low-mass galaxies, but without identified
counterparts in extant optical surveys. Analysis of the resolved
stellar sources in deep $g$- and $i$-band imaging from the WIYN pODI
camera reveals a clustering of possible Red Giant Branch stars
associated with AGC 249525 at a distance of 1.64$\pm$0.45 Mpc.
Matching our optical detection with the \hi\ synthesis map of AGC
249525 from \citet{adams16} shows that the stellar overdensity is
exactly coincident with the highest-density \hi\ contour from that
study. Combining our optical photometry and the \hi\ properties of
this object yields an absolute magnitude of $-7.1 \leq M_V \leq -4.5$,
a stellar mass between $2.2\pm0.6\times10^4~M_{\sun}$ and
$3.6\pm1.0\times10^5~M_{\sun}$, and an \hi\ to stellar mass ratio
between 9 and 144.  This object has stellar properties within the
observed range of gas-poor Ultra-Faint Dwarfs in the Local Group, but
is gas-dominated.
\end{abstract}


\section{Introduction}
\label{sec:intro}
The Arecibo Legacy Fast ALFA survey (ALFALFA) is a blind neutral
hydrogen (\hi) survey carried out with the Arecibo radio telescope
that covers $\sim$7000 square degrees of sky and has cataloged more
than 30,000 sources, the majority of which have never before been
observed in the 21-cm line \citep{giovanelli05, haynes11}.  ALFALFA
detects sources with $2\times10^7$ M$_{\sun}$ of \hi\ at the distance
of the Virgo Cluster and $<10^5$ M$_{\sun}$ within the Local Group at
an angular resolution of $\sim$3.5$'$ and a spectral resolution of
$\sim5$ km~s$^{-1}$ \citep{giovanelli05}.  Among these thousands of
ALFALFA sources is a small population of objects first identified
according to their common characteristics by \citet{giovanelli10} and
dubbed Ultra-Compact High Velocity Clouds (UCHVCs).  The UCHVCs are
very compact ($<10$\arcmin\ diameter), isolated gas clouds that have
\hi\ properties (including narrow velocity widths) that suggest they
may be nearby low-mass galaxies.  The other important characteristic
of UCHVCs is that they have no clear optical counterpart when their
positions and redshifts are checked against optical surveys and
catalogs like the Sloan Digital Sky Survey \citep[SDSS;][]{sdss} and
sources in the NASA Extragalactic Database (NED).

Some have argued that compact high-velocity clouds (CHVCs, with
$\sim$60\arcmin\ diameters) detected in \hi\ surveys are gas clouds in
the Local Group with few or no stars, embedded within low-mass dark
matter halos \citep{blitz99,braun99}.  The detection of a substantial
population of such objects would have implications for the ``missing
satellites problem'', the discrepancy between the numbers of low-mass
halos predicted by $\Lambda$CDM structure formation models compared to
the numbers detected in observational surveys
\citep{kauffmann93,klypin99,moore99}.  However \citet{sternberg02}
demonstrated that the structural properties of the CHVCs do not match
the expected properties of the missing satellites in the $\Lambda$CDM
framework: if extragalactic, they are physically too large.
\citet{giovanelli10} brought renewed attention to this topic by
showing that UCHVCs discovered in ALFALFA, if they are indeed at
distances of $\sim$1 Mpc, {\it do} have sizes and masses consistent
with their being baryonic material embedded within low-mass
($\la10^9$~M$_{\sun}$) dark matter halos, or ``minihalos'', that fit
within the $\Lambda$CDM 
paradigm. \citet{giovanelli10} concluded that the UCHVCs are plausible
minihalo candidates, but exactly what they represent -- e.g., starless
gas clouds, low-mass galaxies with stars that have been missed by past
surveys, or some other poorly-understood HVC phenomenon -- remains
unclear.

A preliminary list of 27 UCHVCs was presented in \citet{giovanelli10},
and \citet{adams13} followed up with a more complete list of 59
objects chosen from the 40\% ALFALFA catalog \citep{haynes11}.  UCHVCs
selected from ALFALFA have \hi\ properties like those of Leo~T
\citep{irwin07}, which is the faintest ($M_V\sim-7$) Local Group dwarf
galaxy with evidence of recent star formation.  At a distance of 1
Mpc, the UCHVCs would have \hi\ masses of $\sim$10$^5$$-$10$^6$ \msun,
\hi\ diameters of $\sim$2$-$3 kpc, and dynamical masses of
$\sim$10$^7$$-$10$^8$ \msun.

In order to derive the masses and sizes of these \hi\ sources, and thus
to better understand their nature, we need to know their distances.
One route for deriving distances for the UCHVCs is to detect their
stellar populations, construct a color-magnitude diagram (CMD), and
then measure a distance from the Tip of the Red Giant Branch (TRGB)
method.  Accordingly, we are carrying out a campaign to obtain deep
optical imaging of UCHVCs with the WIYN 3.5-m telescope\footnote{The
  WIYN Observatory is a joint facility of the University of
  Wisconsin-Madison, Indiana University, the University of Missouri,
  and the National Optical Astronomy Observatory.} to look for stellar
populations associated with these objects, and (if possible) derive
their distances, masses, sizes, and other properties, and to study
their star formation histories.

The first UCHVC we observed with WIYN resulted in the discovery of
Leo~P, an extremely metal-poor gas-rich star-forming dwarf galaxy
\citep{leop1, rhode13, leop3}.  Located just outside the Local Group
at $1.62\pm0.15$ Mpc \citep{mcquinn15}, Leo~P has an \hi\ mass of
$8.1\times10^5$ \msun\ and an \hi-to-stellar mass ratio of $\sim$2.
Its oxygen abundance is $\sim$2\% of the solar value, comparable to
the abundances of I Zw 18, DDO 68, and the ALFALFA galaxy AGC~198691
\citep{alecxy}. Its total magnitude is $M_V=-9.3\pm0.2$ and its
dynamical mass is $1.4\times10^7$ \msun, making Leo P the lowest-mass
galaxy known that is currently forming stars.

Since the discovery of Leo~P, we have developed a systematic observing strategy and procedure for searching for stellar counterparts to the UCHVCs.  Details are presented in \citet{janesh15}, along with a tentative detection of a counterpart to the ALFALFA source AGC~198606, nicknamed “Friend of Leo~T” \citep{adams15} because it is close in position and velocity to dwarf galaxy Leo~T.  We detected a stellar counterpart at a distance of $\sim$380 kpc, with $M_i\sim-4.7$, and an \hi-to-stellar mass ratio of $\sim45-110$. Because our detection has only 92\% significance, additional observations are needed to show unambiguously whether or not this object is a real detection and a new Local Group member.

Other groups are also carrying out optical searches for counterparts
to the ALFALFA UCHVCs \citep[e.g.][]{bellazzini15a, bellazzini15b,
  sand15, beccari16}.  The results from these searches, as well as
from the WIYN imaging obtained thus far, suggest that if nearby
stellar counterparts to UCHVCs can be found, they will much less
obvious than Leo~P. This is not unexpected since ALFALFA sources are,
as mentioned, not classified as UCHVCs unless they have no clear
optical counterpart in extant surveys.

Our campaign to obtain and process WIYN imaging of UCHVCs is underway.
We draw our targets from \citet{adams13} as well as from the 70\%
ALFALFA catalog \citep{jones16}.  In this Letter, we present the
possible detection of a stellar population associated with the UCHVC
AGC~249525, which is included in the 70\% catalog and was
characterized as an excellent dwarf galaxy candidate by
\citet[][hereafter A16]{adams16} based on its \hi\ distribution and evidence of rotation
in follow-up \hi\ synthesis observations.

\section{Observations and Analysis}
\label{sec:methods}

\citetalias{adams16} presented results from Westerbork Radio 
Synthesis Telescope (WRST) observations of several UCHVCs. 
AGC~249525 shows a smooth \hi\ morphology at improved (60\arcsec, 105\arcsec) 
angular resolution and ordered motion consistent with $\sim$15 \kms rotation. 
AGC~249525, and the aforementioned AGC~198606, are among the highest column 
density objects in the UCHVC sample, with peak $N_{HI} \sim 5 \times 10^{19}$ atoms cm$^{-2}$ 
at 105\arcsec~resolution. \citetalias{adams16} reported a mean \hi\ angular 
diameter at half-flux level of 8.5\arcmin\ for AGC~249525 and an \hi\ mass 
of $2.4\times10^6 M_{\sun}$, assuming a distance of 1 Mpc. 
\citetalias{adams16} showed that for a distance range of 0.4$-$2 Mpc, 
and assuming a baryonic mass consisting of only the neutral gas 
component, AGC~249525 falls on the Baryonic Tully-Fisher Relation 
\citep{mcgaugh12} when it is extrapolated to low galaxy masses.

We observed AGC~249525 with the partially-filled One Degree Imager
(pODI) on the WIYN 3.5-m telescope at Kitt Peak National Observatory
on 16 March 2013.
The pODI camera provided a $\sim$24\arcmin~$\times24\arcmin$~field of
view and a pixel scale of 0\farcs11 per pixel. Nine 300s exposures
were obtained in $g$ and $i$ filters. The raw images were transferred
to the ODI Portal, Pipeline, and Archive (ODI-PPA) at Indiana
University and processed using the QuickReduce pipeline
\citep{kotulla14} to remove the instrumental signature. We then
illumination-corrected, scaled, and stacked the images. The FWHM of
the point spread function in the stacked images is 0\farcs78 in $g$
and 0\farcs73 in $i$.

Sources were identified with the IRAF task \texttt{DAOFIND} and the
source lists in $g$ and $i$ were matched. We performed photometry and
calculated final, calibrated $g_0$ and $i_0$ magnitudes for all point
sources in the matched list. Photometric calibration coefficients were
calculated based on SDSS DR13 \citep{dr13} standard stars in the
images and Galactic extinction corrections were calculated using the
relations in \citet{sf11}. The $5\sigma$ detection limit is 25.5 in
$g$ and 24.5 in $i$.

The full details of our analysis methods are given in
\citet{janesh15}. Briefly, we constructed a CMD from the $g_0$ and
$i_0$ photometry and then applied a color-magnitude filter derived
from \citet{girardi} isochrones to the data, sampling a set of
distances between 0.3 and 2.5 Mpc. The CMD filter is intended to
select red giant branch (RGB) stars and horizontal branch (HB) stars
from old, metal-poor stellar populations.  We then smoothed the
spatial distribution of the filtered stars by binning the positions
into a 2-dimensional grid with $\sim8\arcsec\times8\arcsec$ bin sizes,
and convolving the grid with a Gaussian distribution with a
3\arcmin~smoothing radius. Overdensities are regions in the grid that
exceed the mean grid value by some number of standard deviations
($\sigma$).  To quantify the significance of the overdensities, we
created 25000 samples with the same number of points as were found in
the CMD filter, but with a uniform random distribution.  We measured
the peak $\sigma$ value for each of these samples. The distribution is
well-fitted by a lognormal probability distribution function.  To
identify significant overdensities, we calculated the percentage of
the cumulative distribution function (CDF) of peak $\sigma$ values
below that of the overdensity detected in the real data. We consider
an overdensity significant if it has a peak $\sigma$ value higher than
95\% or more of the values in the simulated CDF.

\section{A Possible Optical Counterpart to AGC~249525}
\label{sec:results}
\begin{figure*}[ht!]
\includegraphics[width=\textwidth]{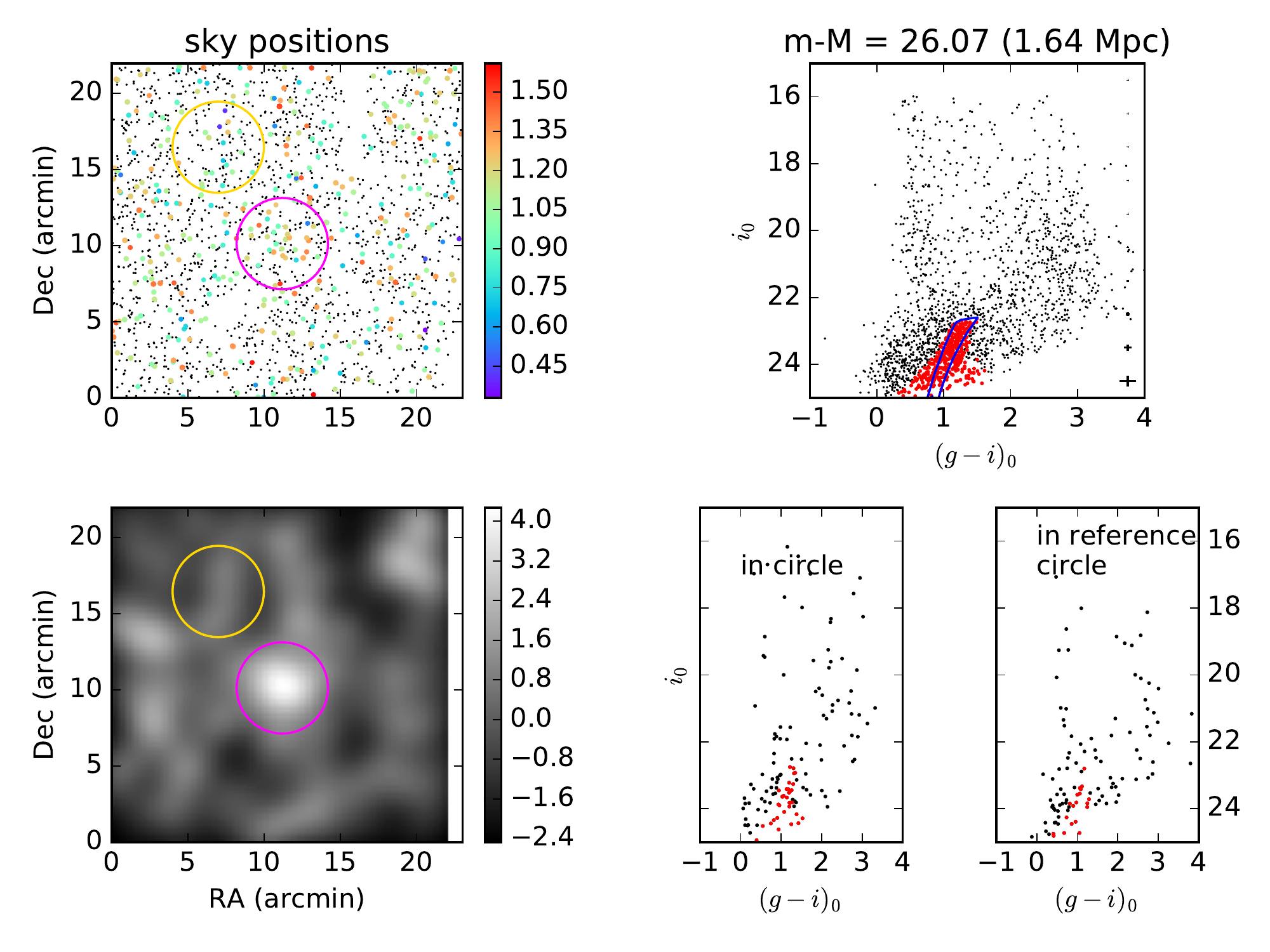}
\caption{Results of the filtering and smoothing process for AGC~249525
  at a filter distance of 1.64 Mpc. Top left: positions of point
  sources relative to the corner of the field. Sources
  that fall outside the CMD filter are marked with small black dots;
  sources within the CMD filter are color-coded
  according to their $(g-i)_0$ values.
A magenta 3\arcmin-radius circle is centered on the detection
peak; a yellow circle of the same size marks a random location used to construct a reference CMD.
Top right: CMD for all point sources in the field; the CMD filter is shown in blue, and the sources selected by the filter are shown in red.
Bottom left: the smoothed stellar density in units
of standard deviations above or below the mean; the 3\arcmin-radius
circle (magenta) is centered on the highest-signal pixel. Bottom
right: CMDs for the stars inside the 3\arcmin-radius magenta circle
(left) and the stars in the yellow reference circle
(right); the latter provides a sampling of the foreground and background contamination present in the detection CMD.\label{fig:detection}}
\end{figure*}

\begin{figure}[ht!]
\includegraphics[width=\columnwidth]{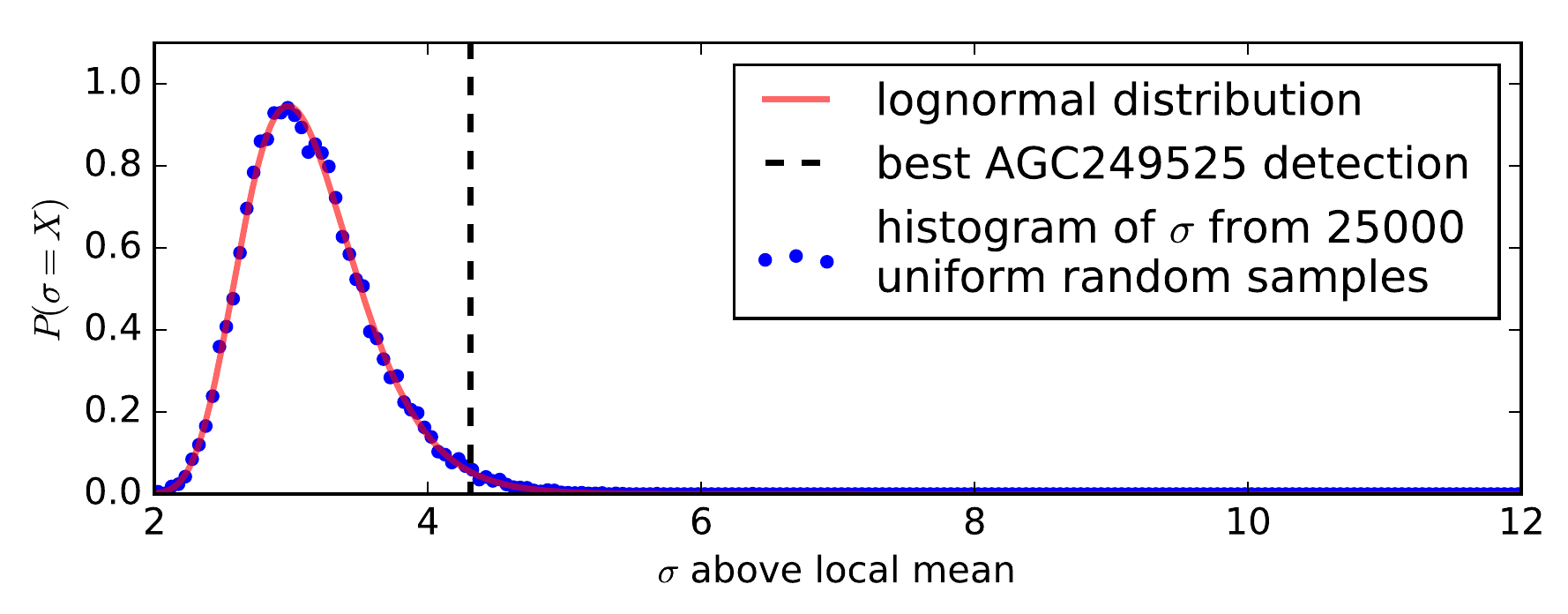}
\caption{Results of the significance testing described in Section
  \ref{sec:methods} for the overdensity shown in Figure
  \ref{fig:detection}. The $\sigma$ value for the overdensity is
  greater than 98.4\% of peak sigma values in 25,000 random
  realizations.  \label{fig:significance}}
\end{figure}

Based on the analysis described above, we found significant stellar
overdensities in the AGC~249525 images at a range of distances between
1.35 Mpc and 2.08 Mpc. In Figure \ref{fig:detection}, we show the
results of the CMD filtering and smoothing process for the
highest-significance detection at 1.64 Mpc $(m-M=26.07\pm0.51)$. We
adopt the distance corresponding to this detection as our assumed
distance to the object. We estimate an uncertainty of $\pm$0.45 Mpc,
based on the observed range of significant ($\geq$95\%) overdensities
that show up at that location.  In Figure \ref{fig:significance}, we
show the results of the significance testing for the detection at 1.64
Mpc. At this distance, 98.4\% of peak overdensities in the random
realizations are weaker than the peak overdensity in the data,
indicating that the overdensity is unlikely to be a random
clustering of sources.

\begin{figure*}[ht!]
\includegraphics[width=\textwidth]{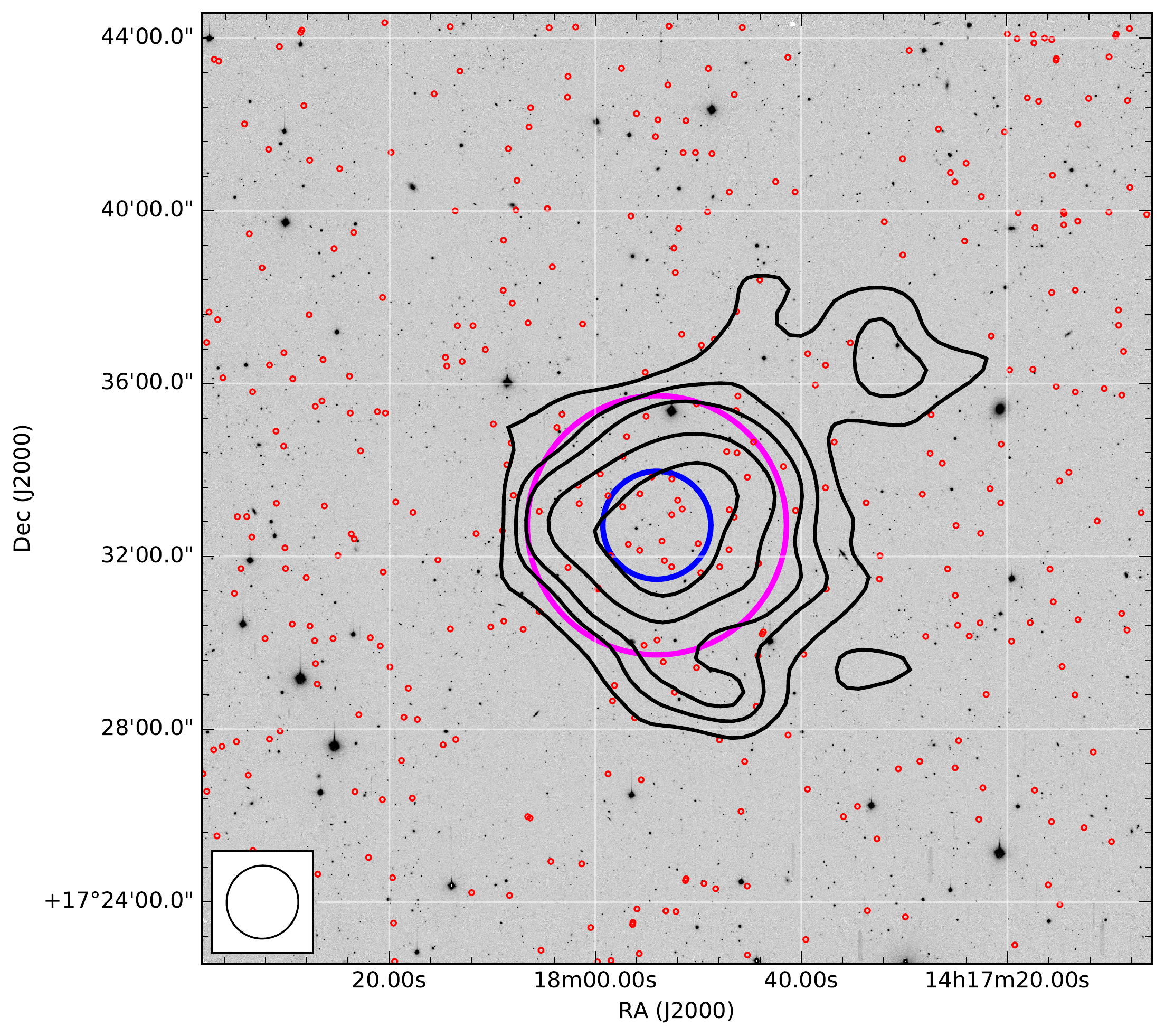}
\caption{The WIYN pODI $i$-band image ($\sim$22\arcmin\ $\times$
  22\arcmin; North-up, East-left) of AGC~249525 with \hi\ contours
  from \citetalias{adams16} overlaid at [9, 15, 20, 30, 40] $\times
  10^{18}$ atoms cm$^{-2}$ (solid black lines). The WSRT beam is shown
  in the bottom-left corner. Also marked are stars selected by the CMD
  filter (red circles) and a 3\arcmin-radius circle at the location of
  the overdensity from Figure \ref{fig:detection} (magenta). A
  75\arcsec-radius circle (blue) shows the aperture used to estimate
  optical properties in Section \ref{sec:props}. The center of the
  stellar overdensity identified by the CMD filtering process
  coincides with the highest column density
  \hi\ contour. \label{fig:contours}}
\end{figure*}

Figure \ref{fig:contours} shows the $i$ image overlaid with the
\hi\ contours from the WSRT data presented in \citetalias{adams16}. The
location of the stellar overdensity, marked by a magenta circle, is
coincident with the highest \hi\ contour level, providing further
evidence that the stellar overdensity is associated with the UCHVC.

\begin{figure*}[ht!]
\includegraphics[width=\textwidth]{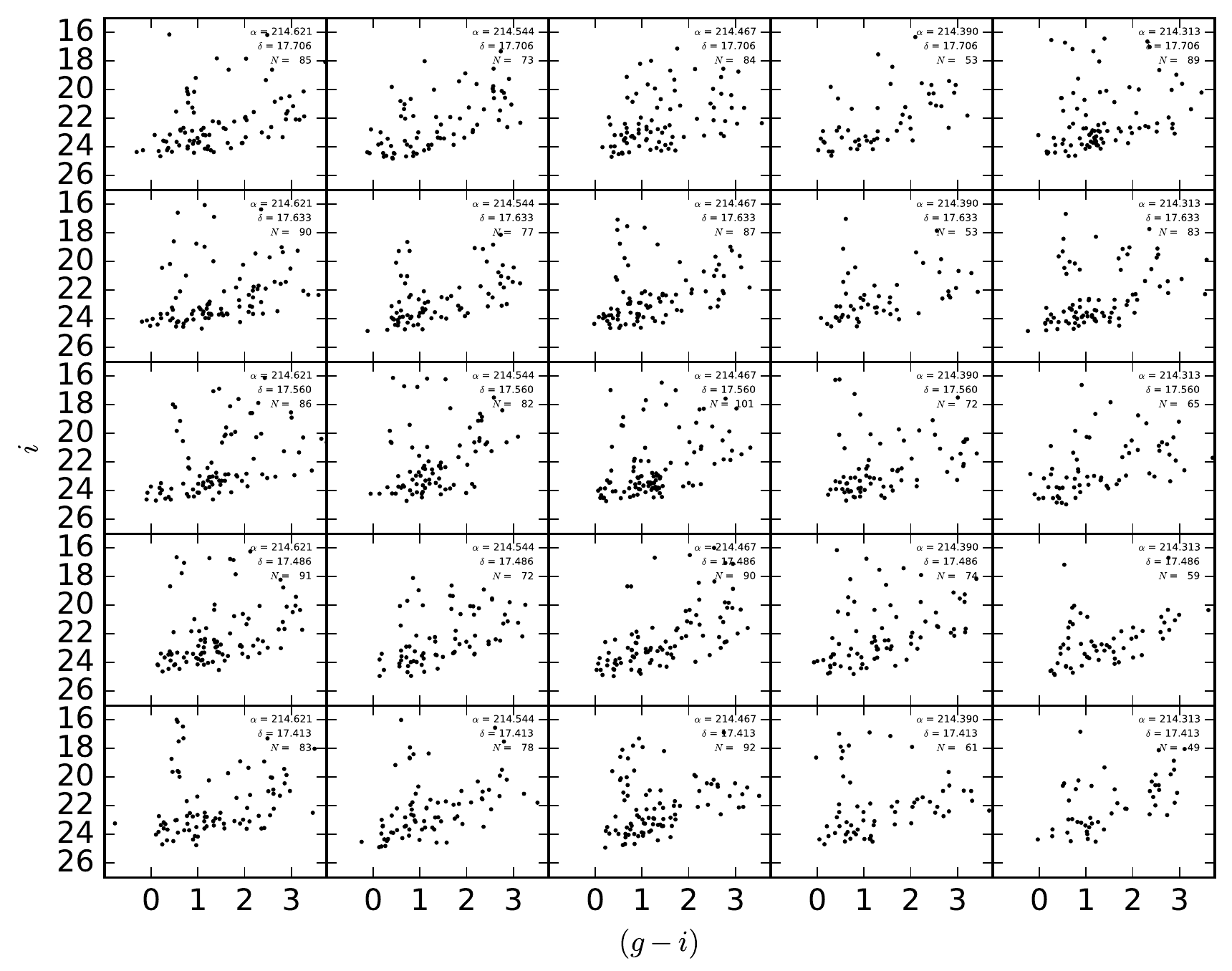}

\caption{CMDs for 4\farcm4 $\times$ 4\farcm4 subsets of the full
  image; the gride is arranged N-up, E-left. 
The central coordinates and the number of stars is printed in the
upper right corner of each box. An overabundance of stars at
$(g-i)_0\simeq1$ appears in the central box, which is coincident with
the overdensity described in Section \ref{sec:results}.
\label{fig:cmdgrid}}
\end{figure*}

Another check on the validity of the overdensity of RGB stars in the
center of the AGC~249525 field is shown in Figure
\ref{fig:cmdgrid}. We divided the image into regions 4\farcm4 $\times$
4\farcm4 in size and plotted the CMD for each portion of the image.
The central region shows an increased number of stars at
$(g-i)_0\simeq1$, consistent with the colors of RGB stars in our CMD
filter. When visualized in this way, the overdensity of potential RGB
stars in the center of the image is readily apparent.  We note that a
modest overdensity appears in the northwest corner of the image (top
right of Figure \ref{fig:cmdgrid}).  The peak $\sigma$ value for this
overdensity is greater than the peak value in only 2\% of the random
realizations, so it is not significant.

\subsection{Estimated \hi\ and Optical Properties}
\label{sec:props}

We can use the 1.64 Mpc distance to calculate the \hi\ properties of
AGC~249525.  Combining the values presented in \citetalias{adams16} with
our distance yields an \hi\ mass of $3.2\pm0.9\times10^6$ $M_{\sun}$,
an \hi\ radius\footnote{Measured at the $1 \times 10^{19}$ atoms
  cm$^{-2}$ level.} of $1.8\pm0.5$ kpc, and a dynamical mass of
$1.6\pm0.5\times10^8$ $M_{\sun}$.

To calculate the observed properties of the optical counterpart, we
masked all objects in the images that are obvious background galaxies
or foreground stars. We then measured the magnitude in the masked
images within a circular aperture of radius 75\arcsec, centered on the
location of the highest-density peak in Figure
\ref{fig:detection}. The 75\arcsec\ radius was chosen to minimize
contributions from sky background fluctuations while maximizing the
number of stars that fell within the CMD filter.
This process yielded 
apparent magnitudes of $g_0=19.61\pm0.10$ and $i_0=18.20\pm0.07$ and a
$(g-i)_0$ color of $1.41\pm0.12$. At a distance of 1.64 Mpc, this
corresponds to an absolute magnitude of $M_V=-7.1$. Because these
values include all unmasked light inside the 75\arcsec~radius
aperture, including fluctuations in the sky background, they should be
considered upper limits.

We also compute lower limit magnitudes by summing the flux from only
those sources selected by the CMD filter that are inside the
75\arcsec-radius circle around the stellar overdensity peak. In this
case we find apparent magnitudes of $g_0=22.10\pm0.02$ and
$i_0=20.93\pm0.02$, with a $(g-i)_0$ color of $1.18\pm0.02$. At a
distance of 1.64 Mpc, the ``minimum'' absolute magnitude of AGC~249525
is $M_V=-4.5$. Although extremely faint compared to most dwarf
galaxies, these absolute magnitudes are consistent with the range of
values for ultra-faint dwarfs (UFDs) in the Local Group and its
environs \citep{mcconnachie12}.

We can use the distance to compute an \hi-to-stellar mass ratio for
this object. We begin by estimating the stellar mass-to-light ratio
using the relations from \citet{bell03}, finding for the redder,
brighter magnitude limit a value of $(M/L)_i=3.79$, and for the bluer,
fainter magnitude limit a value of $(M/L)_i=2.87$. Using these
mass-to-light ratios we calculate a stellar luminosity of
$L_*=9.6\pm2.6\times10^4 L_{\sun}$ and a stellar mass of
$M_*=3.6\pm1.0\times10^5 M_{\sun}$ for the bright limit, and a
luminosity of $L_*=7.7\pm2.1\times10^3 L_{\sun}$ and mass of
$M_*=2.2\pm0.6\times10^4 M_{\sun}$ for the faint limit. Using an
\hi\ mass of $3.2\times10^6 M_{\sun}$, we find \hi-to-stellar mass
ratios of $M_{\hi}/M_*\sim9$ for the bright limit and
$M_{\hi}/M_*\sim144$ for the faint limit. The \hi\ and 
optical properties are listed in Table \ref{tab:props}.

\citetalias{adams16} discuss the potential close neighbors of AGC~249525: Bootes I, Bootes II, and UGC~9128. Bootes I and Bootes II are both located within 60 kpc of the Sun, and so are not in close spatial proximity to AGC~249525. At the distance we have estimated here, the most likely neighbor is the galaxy UGC~9128 at 2.27 Mpc \citep{tully13}.  UGC~9128 has a $cz=152$ \kms \citep{mcconnachie12} and is less than 10\arcdeg~away from AGC~249525, which has a $cz=48$ \kms.

\begin{deluxetable}{lll}
\tablecaption{Properties of AGC249525 \label{tab:props}}
\tablecolumns{3}
\tablenum{1}
\tablewidth{\columnwidth}
\tablehead{
\colhead{Property} &
\nocolhead{Value} &
\colhead{Value}
}
\startdata
RA (J2000)        & \multicolumn{2}{r}{$14^{\mathrm{h}}~17^{\mathrm{m}}~53\fs4$ } \\
Dec (J2000)       & \multicolumn{2}{r}{17\arcdeg\ 32\arcmin\ 42\farcs3 } \\
Distance          & \multicolumn{2}{r}{$1.64\pm0.45$ Mpc} \\
$cz$              & \multicolumn{2}{r}{47 \kms} \\
\hi\ mass         & \multicolumn{2}{r}{$3.2\pm0.9\times10^6 M_{\sun}$} \\
\hi\ radius       & \multicolumn{2}{r}{$1.8\pm 0.5$ kpc} \\
Rotational velocity & \multicolumn{2}{r}{$15_{-2}^{+6}$ \kms} \\
Dynamical mass    & \multicolumn{2}{r}{$1.6\pm0.5\times10^8 M_{\sun}$} \\
\tableline
                  & Lower Limit & Upper Limit \\
\tableline
$i_0$             & $20.9\pm0.02$ & $18.20\pm0.07$ \\
$M_V$             & -4.5 & -7.1 \\
$(g-i)_0$         & $1.18\pm0.02$ & $1.41\pm0.12$ \\
$L_*$             & $7.7\pm2.1\times10^3 L_{\sun}$ & $9.6\pm2.6\times10^4 L_{\sun}$ \\
Stellar mass      & $2.2\pm0.6\times10^4 M_{\sun}$ & $3.6\pm1.0\times10^5 M_{\sun}$ \\
$M_{\hi}/M_*$     & 144 & 9 \\
\enddata
\end{deluxetable}

\section{AGC~249525: A Gas-rich Ultra-faint Dwarf Galaxy?}
\label{sec:discussion}
The possible detection of an optical counterpart for the UCHVC
AGC~249525 has interesting implications. Along with AGC198606
\citep[][]{janesh15}, it would represent one of the most extreme
galaxies in or near the Local Group, with its sparse stellar
population and large $M_{\hi}/M_*$ ratio. While galaxies like these
are difficult to detect, models have predicted that they should
exist. High-resolution hydrodynamic cosmological simulations from
\citet{onorbe15} indicate that isolated, low-mass dark matter halos
result in galaxies with stellar masses between $10^4-10^6 M_{\sun}$
and large $M_{gas}/M_*$ ratios and stellar masses between $10^4-10^6
M_{\sun}$, consistent with the range of possible values for
AGC~249525. Additionally, the N-body and semi-analytic models of
\citet{br11} predict that isolated UFDs should exist in significant
numbers in the Local Volume, with optical properties similar to known
UFDs, though most of these predicted UFDs have lost their gas via
stripping or stellar feedback effects.

The putative optical counterpart for AGC~249525 has an absolute
magnitude that places it in the middle of the range of the faint dwarf
galaxies in or near the Local Group. UFDs like Segue~1 ($M_V=-1.5$),
Segue~2 ($M_V=-2.5$), and Willman~1 ($M_V=-2.7$) are at the faintest
end of the range of known nearby dwarfs
\citep{mcconnachie12}. \citet{koposov15} recently discovered nine new
UFDs in the Southern Hemisphere with $M_V=-2.0$ to $-$6.6. The
gas-rich dwarfs Leo~P \citep[$M_V = -9.3$;][]{rhode13, mcquinn15} and
Leo~T \citep[$M_V = -8.0$;][]{irwin07} are significantly brighter,
though they are actively forming, or have recently formed, stars. The
possible optical counterpart to AGC~198606, which is gas-rich but
without active star formation, is closer to the UFDs in its
luminosity, with $M_V\sim-3.5$. AGC~249525, while more luminous than
AGC~198606, is still substantially fainter than Leo T and Leo P.

Previous work has concluded that some UFDs could be the stellar
populations left over after their progenitor dwarf galaxies were
stripped of gas and stars by the Milky Way \citep{willman06,
  martin07}. In \citet{janesh15} we speculated that, with the
discovery of gas-rich galaxies with extremely sparse stellar
populations like AGC~198606 and now AGC~249525, a different formation
scenario for UFDs is possible. With less massive progenitors, less
stripping would be necessary to reach the low stellar masses and faint
total magnitudes observed in UFDs.

Further observations are needed to confirm the optical counterpart to
AGC~249525. Only the upper portion of the RGB is accessible in the
current WIYN pODI data, whereas a detection of HB stars would provide
a more definitive distance determination. Our derived distance is
similar to that of Leo P. \citet{mcquinn15} used Hubble Space
Telescope (HST) observations to measure a distance to Leo~P of
$1.62\pm0.15$ Mpc, based on a combination of TRGB stars, HB stars, and
RR Lyrae variables. At 1.6 Mpc, HB stars would have $M_V\sim26.5$, so
a robust detection of the HB in AGC~249525 
would therefore likely require very deep imaging with a large-aperture
ground-based telescope or HST. Spectroscopic radial velocity
measurements of the putative RGB stars in this object would also help
confirm the association of stars with the \hi\ gas, although at these
magnitudes such observations would be a challenge.

\acknowledgments 
We thank the anonymous referee for helpful comments.
We thank the WIYN, NOAO, and ODI-PPA staff for their help at various stages
of this project.  W.F.J. and K.L.R. are supported by NSF grant
AST-1615483.  S.J. acknowledges support from the Australian Research
Council's Discovery Project funding scheme (DP150101734).  EAKA is
supported by TOP1EW.14.105, which is financed by the Netherlands
Organisation for Scientific Research (NWO).  The ALFALFA team at
Cornell is supported by NSF grants AST-0607007 and AST-1107390 to
R.G. and M.P.H. and by grants from the Brinson Foundation.  J.M.C. is
supported by NSF grant 1211683.  This research made use of the
NASA/IPAC Extragalactic Database (NED) which is operated by the Jet
Propulsion Laboratory, California Institute of Technology, under
contract with the National Aeronautics and Space Administration.

\clearpage

\end{document}